\documentclass[twocolumn,english,prl,floatfix,twocolumn,showpacs]{revtex4}
\usepackage[T1]{fontenc}
\usepackage[latin9]{inputenc}
\usepackage[active]{srcltx}
\usepackage{color}
\usepackage{amstext}
\usepackage{amssymb}
\usepackage{graphicx}
\usepackage{esint}

\makeatletter
\@ifundefined{textcolor}{}
{%
 \definecolor{BLACK}{gray}{0}
 \definecolor{WHITE}{gray}{1}
 \definecolor{RED}{rgb}{1,0,0}
 \definecolor{GREEN}{rgb}{0,1,0}
 \definecolor{BLUE}{rgb}{0,0,1}
 \definecolor{CYAN}{cmyk}{1,0,0,0}
 \definecolor{MAGENTA}{cmyk}{0,1,0,0}
 \definecolor{YELLOW}{cmyk}{0,0,1,0}
 }

\global\long\def\ket#1{\left| #1\right\rangle }

\global\long\def\bra#1{\left\langle #1 \right|}

\global\long\def\kket#1{\left\Vert #1\right\rangle }

\global\long\def\bbra#1{\left\langle #1\right\Vert }

\global\long\def\braket#1#2{\left\langle #1\right. \left| #2 \right\rangle }

\global\long\def\bbrakket#1#2{\left\langle #1\right. \left\Vert #2\right\rangle }

\global\long\def\av#1{\left\langle #1 \right\rangle }

\global\long\def\tr{\text{Tr}}

\global\long\def\im{\text{Im}}

\global\long\def\re{\text{Re}}

\global\long\def\sign{\text{sign}}

\newcommand{\be}{\begin{eqnarray}}\newcommand{\ee}{\end{eqnarray}}\def\beq{\begin{equation}}\def\eeq{\end{equation}}

\usepackage{babel}

\usepackage{babel}

\makeatother

\usepackage{babel}
\begin{document}

\title{Combined effect of thermal and quantum fluctuations in superconducting
nanostructures: a path integral approach }

\author{Pedro Ribeiro}

\affiliation{Max Planck Institute for the Physics of Complex Systems, Nothnitzer
Str. 38, 01187 Dresden, Germany}

\author{Antonio M. García-García}

\affiliation{University of Cambridge, Cavendish Laboratory, JJ Thomson Avenue,
Cambridge, CB3 0HE, UK}
\begin{abstract}
We study the combined effect of thermal and quantum fluctuations in
a zero dimensional superconductor. By using path integral techniques,
we obtain novel expressions for the partition function and the superconducting
order parameter which include both types of fluctuations. Our results
are valid for any temperature and to leading order in $\delta/\Delta_{0}$
where $\delta$ is the mean level spacing and $\Delta_{0}$ is the
bulk energy gap. We avoid divergences at low temperatures, previously
reported in the literature, by identifying and treating non-perturbatively
a low-energy collective mode. In the low and high temperature limit
our results agrees with those from the random phase (RPA) and the
static path approximation (SPA) respectively. 
\end{abstract}


\maketitle
\global\long\def\ket#1{\left| #1\right\rangle }

\global\long\def\bra#1{\left\langle #1 \right|}

\global\long\def\kket#1{\left\Vert #1\right\rangle }

\global\long\def\bbra#1{\left\langle #1\right\Vert }

\global\long\def\braket#1#2{\left\langle #1\right. \left| #2 \right\rangle }

\global\long\def\bbrakket#1#2{\left\langle #1\right. \left\Vert #2\right\rangle }

\global\long\def\av#1{\left\langle #1 \right\rangle }

\global\long\def\tr{\text{Tr}}

\global\long\def\im{\text{Im}}

\global\long\def\re{\text{Re}}

\global\long\def\sign{\text{sign}}

\global\long\def\Det{\text{Det}}

Superconductivity in nano-structures has attracted the attention of
theorists \cite{anderson} and experimentalists \cite{giaver} since
the early days of the Bardeen-Cooper-Schriffer (BCS) theory. Explicit
calculations \cite{bat} soon showed that, even within a mean field
formalism, finite size effects had a profound impact on the superconducting
state. Experimentally it was also observed \cite{rt,burham} that
the superconducting transition in nanowires and small particles became
broader as the grain size decreases due to thermal and quantum fluctuations.
The use of path integral techniques \cite{scalapino,halperin} led
to a quantitative description of thermal fluctuations specially in
zero dimensional superconductors \cite{scalapino} where the SPA is
applicable. By contrast semi-phenomenological models \cite{giordano}
that combined quantum and thermal fluctuations provided only a qualitative
description of the broadening of the transition observed in superconducting
nanowires. The field received an important impetus in the mid nineties
after the experiments of Ralph et al. on single, isolated Al nanoparticles
\cite{tinkham} that showed for the first time that superconductivity
survived in single particles down to a few nanometers. These experiments
also stimulated the theoretical interest in ultrasmall superconductors.
At zero temperature, the Richardson's formalism \cite{richardson},
originally introduced in the context of nuclear physics, made possible
to find exact solutions for the low energy excitations of the reduced
BCS Hamiltonian \cite{vondelft}. However a theoretical analysis that
takes into account thermal and quantum fluctuations simultaneously
is still an open problem in the field. In \cite{argent} this problem
was addressed by combining the SPA, that models thermal fluctuations,
with the RPA, that accounts for quantum fluctuations to leading order
in $\delta/\Delta_{0}$. However it was found that the resulting partition
function had singularities at low temperature. Progress in this problem
are specially timely as recent experiments, taking advantage of advances
in the growth and control of nanostructures, have put the basis to
test quantitatively the limits of superconductivity in the nanoscale
\cite{pbnew,alu,ultra2d,nmat,zgirski}. 

The main goal of this paper is to put forward a theoretical analysis
free of divergences and valid at all temperatures that combines thermal
fluctuations and quantum fluctuations to leading order in $\delta/\Delta_{0}$
in a zero dimension superconductor. These results are of interest
for other strongly interacting Fermi systems beyond the realm of superconductivity
in nanostructures. Typical examples include hot nuclei (see \cite{nuclear}
and references therein), and trapped cold atomic gases \cite{heiselberg}.
Technical details of the calculation are postponed to a forthcoming
publication \cite{pedro2}. Here we summarize the main results, their
limits of applicability and some technical aspects. Moreover we also
provide explicit results of the order parameter in two simple cases:
a constant spectral density and a single, highly degenerate, shell.\textit{}\\
\textit{}\\
\textit{Model, approximations and main results }\\
We consider the BCS Hamiltonian: 
\begin{eqnarray}
H & = & \sum_{\alpha,\sigma}\varepsilon_{\alpha}\, c_{\alpha,\sigma}^{\dagger}c_{\alpha,\sigma}\label{eq:Hamilton}\\
 &  & -\delta g\left(\sum_{\alpha,\alpha'}I(\alpha,\alpha')c_{\alpha,1}^{\dagger}c_{-\alpha,-1}^{\dagger}c_{-\alpha',-1}c_{\alpha',1}\right).\nonumber 
\end{eqnarray}
where $\alpha,-\alpha$ label one-particle states related by time
reversal symmetry with energies $\varepsilon_{\alpha}=\varepsilon_{-\alpha}$,
$\delta$ is the mean level spacing, $\sigma=\pm1$ is the spin label,
$g$ the dimensionless coupling constant, $I(\alpha,\alpha')=L^{d}\int d^{d}r\Psi_{\alpha}^{2}(r)\Psi_{\alpha'}^{2}(r)$
with $\Psi_{\alpha}$ an eigenstate of the one-body problem and $L$
the system size. The partition function of the model is given by $Z=\tr\left[e^{-\beta H}\right]$.
Fermionic degree of freedom can be integrated exactly by introducing
a complex valued Hubbard-Stratonovich field $\Delta(\tau,r)$ which
results in a partition function, 
\begin{eqnarray}
\frac{Z}{Z_{0}} & = & \int\mathcal{D}\Delta^{\dagger}\mathcal{D}\Delta\ e^{-S\left[\Delta\right]}\label{eq:parti}
\end{eqnarray}
where $Z_{0}$ is the partition function for free electrons. The main
goal of the paper is to evaluate (\ref{eq:parti}), including thermal
and quantum fluctuations. The main approximations in our calculation
are: a) the grain size is zero dimensional, namely, the coherence
length $\xi$, is smaller than the system size. As a consequence,
$\Delta(\tau,r)$ only depends on imaginary time $\Delta(\tau,r)\approx\Delta(\tau)$;
b) the time dependence is sufficiently weak so that an expansion to
second order is justified. At $T=0$ this corresponds with the usual
RPA around the saddle point solution $\Delta_{0}$ which is valid
in the limit $\delta/\Delta_{0}<1$; c)\textcolor{red}{{} }\textcolor{black}{the
interacting region in the BCS Hamiltonian is restricted to the a narrow
interval $[-E_{D},E_{D}]$ around the chemical potential where $E_{D}$
is the Debye energy. We assume that $I(\alpha,\alpha')=L^{d}\int d^{d}r\Psi_{\alpha}^{2}(r)\Psi_{\alpha'}^{2}(r)\approx1$
is independent on $\alpha,\alpha'$ and $I(\alpha,\alpha')=0$ outside
the interacting region. }This is a good approximation as it was shown
in \cite{usprl} that the $\alpha$ dependence is only important for
$\xi<L$ ; d) we assume that for $\delta/\Delta_{0}<1$ Coulomb interactions
can be accounted by a simple redefinition of $g$. Recent experiments
\cite{nmat} suggest that, at least for Pb and Sn grains, this is
a good approximation up to sizes $\delta\sim\Delta_{0}$ or $L\sim5$nm. 

We also note that, for the sake of simplicity, no observable sensitive
to odd-even effects are considered. However, even in a grand canonical
formalism, it is possible to take them into account \cite{larkin},
at least for low temperatures, by simply blocking the level closest
to the Fermi energy. 

The main result of this letter is the following expression for $Z$,
valid at all temperatures, that includes simultaneously thermal and
quantum fluctuations,
\begin{eqnarray}
\frac{Z}{Z_{0}} & = & \int_{0}^{\infty}ds_{0}^{2}\ e^{-\beta\left(\mathcal{A}_{0}\left[s_{0}\right]+\mathcal{A}_{1}\left[s_{0}\right]\right)}\label{eq:parmain}
\end{eqnarray}
where
\begin{eqnarray}
\mathcal{A}_{0}\left[s_{0}\right] & = & \left(\delta g\right)^{-1}s_{0}^{2}\\
 &  & -\frac{2}{\beta}\int_{D}d\varepsilon\varrho\left(\varepsilon\right)\ln\left[\frac{\cosh\left(\frac{\beta\xi}{2}\right)}{\cosh\left(\frac{\beta|\varepsilon|}{2}\right)}\right],\nonumber \\
\mathcal{A}_{1}\left[s_{0}\right] & = & \frac{1}{2}\int d\nu\,\left[n_{b}\left(\nu\right)-\frac{1}{\beta\nu}\right]\\
 &  & \times\frac{\ln\left[\widetilde{C}\left(\nu+i0^{+}\right)\right]-\ln\left[\widetilde{C}\left(\nu-i0^{+}\right)\right]}{2\pi i},\nonumber 
\end{eqnarray}

$\xi=\sqrt{s_{0}^{2}+\varepsilon^{2}}$, $\varrho\left(\varepsilon\right)=\sum_{\alpha}\delta\left(\varepsilon-\varepsilon_{\alpha}\right)$
is the spectral density of the one-body problem, $n_{f}\left(z\right)=\frac{1}{e^{\beta z}+1},\, n_{b}\left(z\right)=\frac{1}{e^{\beta z}-1}$
are the Fermi and Bose function respectively and

\begin{eqnarray*}
\widetilde{C}\left(z\right) & = & \left(-z^{2}+4s_{0}^{2}\right)\left(-z^{2}\right)\left[\int_{D}d\varepsilon\,\varrho\left(\varepsilon\right)\frac{r\left(\xi\right)}{-z^{2}+\left(2\xi\right)^{2}}\right]^{2}\\
 &  & +\left(-z^{2}\right)\left[\int_{D}d\varepsilon\,\varrho\left(\varepsilon\right)\frac{2\varepsilon\, r\left(\xi\right)}{-z^{2}+\left(2\xi\right)^{2}}\right]^{2},
\end{eqnarray*}
with $r\left(\xi\right)=\frac{1}{2\xi}\tanh\left(\frac{\beta\xi}{2}\right)$
and\textcolor{black}{{} $\int_{D}=\int_{-E_{D}}^{E_{D}}$.} For $T=0$
we recover the RPA results \cite{argent} and, for $T\gg T_{c}$,
$Z$ is given by the SPA of \cite{scalapino}.\textit{}\\
\textit{}\\
\textit{Calculation Highlights}\\
We give an overview of the calculation leading to (\ref{eq:parmain})
with special emphasis on the main differences with respect to the
techniques of \cite{argent}. A detailed account of technical details
will be provided elsewhere \cite{pedro2}. 

The task is to evaluate \textit{simultaneously} the contribution to
the partition function $Z$ of thermal fluctuations, taken into account
by integrating exactly over the static component of $\Delta(\tau)$
(SPA) and quantum fluctuations, arising as small (imaginary) time
dependent Gaussian corrections (RPA) to SPA. Previous approaches to
this problem \cite{argent} considered indeed small corrections to
a static solution $\Delta(0)$, $\Delta\left(\tau\right)=\Delta(0)+\delta\Delta\left(\tau\right)$
where $\delta\Delta\left(\tau\right)$ is the second derivative matrix
for the fluctuations $\delta\Delta\left(\tau\right)$. Then $\Delta(0)$
is integrated out exactly but the integral over $\delta\Delta\left(\tau\right)$
is carried out in the Gaussian approximation only. It is therefore
assumed that any small correction around any $\Delta(0)$ is still
a local minimum of the action, namely, the real part of the eigenvalues
of $\delta\Delta\left(\tau\right)$ is always positive. However it
was found in \cite{argent} that some eigenvalues of $\delta\Delta\left(\tau\right)$
acquire a negative real part as the temperature is lowered. As a consequence
divergences occur and the theory breaks down, preventing thus the
combined study of quantum and thermal fluctuations. Divergences in
this context usually suggest the existence of a collective zero mode
that must be treated non-perturbately. 

In order to identify this collective mode we separate phase and amplitude
fluctuations by using polar coordinates $\Delta(\tau)=s(\tau)e^{i\phi(\tau)}$,
with $s(\tau)=s_{0}+\delta s(\tau)$ and $\phi(\tau)=a_{0}\tau+\phi_{0}+\delta\phi(\tau)$,
where $\delta\phi(\tau)$ and $\delta s(\tau)$ are small fluctuations
around the static values $s_{0},\phi_{0}$ and $a_{0}=\frac{2\pi}{\beta}M\ \left(M\in\mathbb{Z}\right)$
accounts for phase configurations with non-trivial winding numbers.
Only the $a_{0}=0$ is considered, contributions from different values
of $a_{0}$, known to be related to odd-even effects \cite{larkin}
are not addressed here. The identification of this collective mode
is nevertheless crucial, if treated perturbatively it will lead to
the negative eigenvalues and divergences observed in \cite{argent}.
By contrast, following the above decomposition, the eigenvalues of
$\delta\phi(\tau)$ and $\delta s(\tau)$ in our case have always
a positive real part and therefore no divergences arise. We can then
treat separately the collective mode and integrate exactly over the
static phase $\phi_{0}$. This is the key difference between our method
and that of \cite{argent}. As a consequence our results provide a
quantitative, free of divergences, description of the combined quantum
and thermal fluctuations at any temperature.\textit{}\\
\textit{}\\
\textit{Results}\\
The natural order parameter for the superconducting transition is
the connected pair correlation function $\Delta_{C}^{2}=\left(g\delta\right)^{2}\sum_{\alpha\alpha'}\av{c_{\alpha'1}^{\dagger}c_{\alpha'-1}^{\dagger}c_{\alpha-1}c_{\alpha1}}_{C}$.
An explicit expression for $\Delta_{C}$ is obtained in a standard
way by adding source terms to the action (\ref{eq:parti}) and deriving
with respect to them, 
\begin{eqnarray}
\Delta_{C}^{2} & = & \bar{\Delta}^{2}\label{eq:Gap_C}\\
 &  & -\left(\delta g\right)^{2}\int_{D}d\varepsilon\varrho\left(\varepsilon\right)\left[\av{\av{n_{\text{sc}}\left(\xi\right)^{2}}}-\av{\av{n_{\text{sc}}\left(\xi\right)}}^{2}\right],\nonumber 
\end{eqnarray}
where 
\begin{eqnarray}
\bar{\Delta}^{2} & = & \av{\av{s_{0}^{2}\left[\left(\delta g\right)\int_{D}d\varepsilon\varrho\left(\varepsilon\right)\, r\left(\xi\right)\right]^{2}}},\label{eq:bardel}
\end{eqnarray}
$n_{\text{sc}}\left(\xi\right)=\frac{1}{2}\left[1-\frac{\varepsilon}{\xi}\tanh\left(\frac{\beta\xi}{2}\right)\right]$,
and the average $\av{\av{...}}$ is defined as

\begin{equation}
\av{\av O}=\frac{Z_{0}}{Z}\int_{0}^{\infty}ds_{0}^{2}\ e^{-\beta\left(\mathcal{A}_{0}\left[s_{0}\right]+\mathcal{A}_{1}\left[s_{0}\right]\right)}\, O.
\end{equation}
In the literature other parameters have been considered to study deviations
from mean-field results: for example $\av{\av{s_{0}^{2}}}$ \cite{scalapino}
and $\Delta_{P}^{2}=\left(g\delta\right)^{2}\sum_{\alpha\alpha'}\left[\av{c_{\alpha'1}^{\dagger}c_{\alpha'-1}^{\dagger}c_{\alpha-1}c_{\alpha1}}_{g}-\av{c_{\alpha'1}^{\dagger}c_{\alpha'-1}^{\dagger}c_{\alpha-1}c_{\alpha1}}_{g=0}\right]$
\cite{argent}. The latter can be simply related to (\ref{eq:bardel})
by $\Delta_{P}^{2}=\bar{\Delta}^{2}-g\delta\left(\delta g\right)\int_{D}d\varepsilon\varrho\left(\varepsilon\right)\left[\av{\av{n_{\text{sc}}\left(\xi\right)^{2}}}-n_{f}\left(\varepsilon\right)^{2}\right]$.
For simplicity we assume $\Delta_{C}\approx\bar{\Delta}$ as other
terms in (\ref{eq:Gap_C}) do not play a significant role and make
the calculation slightly more involved. $\Delta_{C}$ becomes the
bulk gap for $\delta\to0$, and it is expected to be closely related
to the spectral gap at finite $\delta.$ We focus on two specially
simple situations: a) a constant spectral density, b) only one level,
usually called shell, in the interacting region with a degeneracy
$N_{l}\gg1$ such that $\delta/\Delta_{0}\ll1$ where $\delta=2E_{D}/N_{l}$.
Physically this corresponds to a spherical or cubic grain in which,
due to geometrical symmetries, the spectrum is highly degenerate.
Other geometries can be easily studied but calculations are more involved.
We postpone this study to a future publication \cite{pedro2}.\textit{}\\
\textit{}\\

\begin{figure}
\centering{}\includegraphics[width=7.2cm,height=5cm,keepaspectratio]{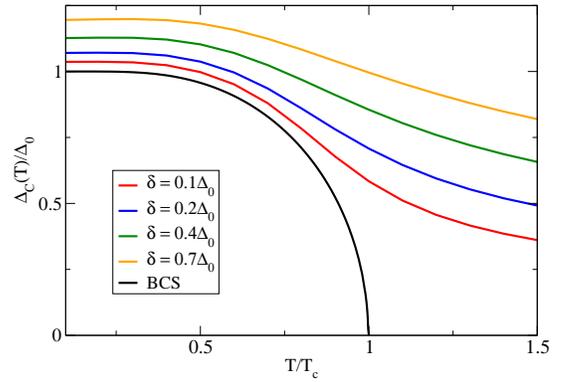}\caption{\label{fig:1} $\Delta_{C}(T)$, (\ref{eq:Gap_C}), for a constant
spectral density $\rho(\varepsilon)=1/\delta$. $\Delta_{C}(T)$ combines
thermal and quantum fluctuations. It reduces to the RPA (SPA) for
$T\ll T_{C}$$(T\gg T_{c})$ . }
\end{figure}
\begin{figure}
\centering{}\includegraphics[width=7.2cm,height=5cm,keepaspectratio]{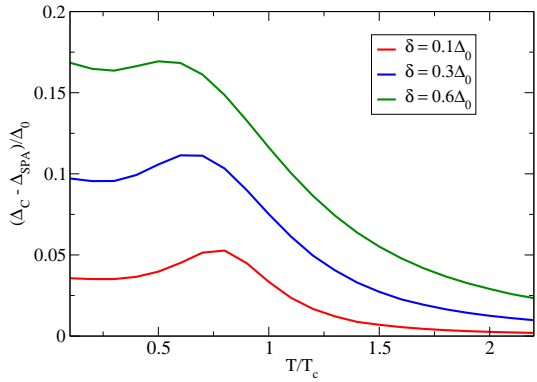}\caption{\label{fig:2} Difference between $\Delta_{C}(T)$ (\ref{eq:Gap_C})
and the SPA prediction \cite{scalapino} which only takes into account
thermal fluctuations for different $\delta$'s. It is assumed that
$\rho(\epsilon)\approx1/\delta$. For low temperatures the difference
is just the usual RPA correction that describes quantum fluctuations
at $T=0$. However the peak observed close to $T_{c}$ is due to the
non-trivial interplay of thermal and quantum fluctuations which is
beyond the reach of SPA and RPA separately. This is the first time
that this region is accessible to analytical techniques. }
\end{figure}

\textit{Constant spectral density}\\
In this case $\varrho\left(\varepsilon\right)=1/\delta$ and the partition
function (\ref{eq:parti}) cannot be simplified further so we carry
out the calculation of $\Delta_{C}$ (\ref{eq:Gap_C}) numerically.
In Fig.\ref{fig:1} we depict $\Delta_{C}(T)$ for different values
of $\delta$$.$ As was expected, no divergences arise at low temperatures.
For zero temperature $\Delta_{C}(0)$ is equal to the RPA result \cite{argent}
that predicts a leading correction $\Delta_{C}(0)=\Delta_{0}(1+\alpha\delta/E_{D})$
with $\alpha$ a constant of order the unity. For $T\gg T_{C}$ ,
$\Delta_{C}$ agrees with the SPA \cite{scalapino} that describes
thermal but not quantum fluctuations (see Fig. \ref{fig:2}). Results
from Richardson's formalism \cite{richardson,altshu} at $T=0$ are
similar but a direct comparison is not possible as $\Delta_{C}$ is
not exactly the spectral gap. In Fig. \ref{fig:2} we depict the difference
between (\ref{eq:bardel}) and the SPA prediction. Deviations at low
temperatures are mostly due to the RPA correction, however it is clearly
observed that, for intermediate temperatures, differences from SPA
results increase as a consequence of the combined effect of thermal
and quantum fluctuations. Previously this region was not accessible
to analytical calculations. We note that the observed enhancement
of $\Delta_{C}$ by quantum and thermal fluctuations is not an indication
that superconductivity is more robust. In fact fluctuations always
weaken long range order causing phase slips and the broadening of
the transition. The gap is enhanced because fluctuations induce pairing
in circumstances which are not allowed by a mean field formalism.\textit{}\\
\textit{}\\
\textit{Shell models}\\
The calculation of the partition function greatly simplifies by assuming
that there are only two degenerate levels (shells) in the interaction
region. We note that quantum fluctuations are still small, and therefore
our formalism is still applicable, provided that the degeneracy of
the level $N_{l}/2$ is large enough such that $\delta\ll\Delta_{0}$
where in this case $\delta=2E_{D}/N_{l}$. With this simplification
it is possible to find an explicit expression for $\mathcal{A}_{1}$.
For two shells with energy at $\pm\varepsilon_{0}$ (\textit{i.e.}
$\varrho\left(\varepsilon\right)=\frac{N_{l}}{2}\left[\delta\left(\varepsilon-\varepsilon_{0}\right)+\delta\left(\varepsilon+\varepsilon_{0}\right)\right]$)
$\mathcal{A}_{0},\mathcal{A}_{1}$ in (\ref{eq:parmain}) are given
by,
\begin{eqnarray*}
\mathcal{A}_{0}\left[s_{0}\right] & = & \delta g\left\{ s_{0}^{2}-\frac{4\varepsilon_{0}\coth\left(\frac{\varepsilon_{0}\beta_{c}}{2}\right)\log\left[\frac{\cosh\left(\frac{1}{2}\beta\sqrt{\varepsilon_{0}^{2}+s_{0}^{2}}\right)}{\text{sech}\left(\frac{\beta|\varepsilon_{0}|}{2}\right)}\right]}{\beta}\right\} \\
\mathcal{A}_{1}\left[s_{0}\right] & = & \frac{1}{\beta}\ln\left[\frac{\beta\xi_{0}^{2}\text{csch}^{2}\left(\beta\xi_{0}\right)\sinh\left(\beta s_{0}\right)}{s_{0}}\right]
\end{eqnarray*}
where $\xi_{0}=\sqrt{s_{0}^{2}+\varepsilon_{0}^{2}}$ and $\beta_{c}=T_{c}^{-1}=\frac{2\coth^{-1}\left(\frac{E_{D}g}{\varepsilon_{0}}\right)}{\varepsilon_{0}}$.
For $T=0$ the first correction to the mean-field result coincides
with the RPA prediction, $\Delta_{C}=\Delta_{0}\left[1+\frac{g\delta}{\Delta_{0}}\left\{ \sqrt{1+\left(\frac{\varepsilon_{0}}{\Delta_{0}}\right)^{2}}-\frac{1}{2}\left[1+\left(\frac{\varepsilon_{0}}{\Delta_{0}}\right)^{2}\right]\right\} \right]$,
where $\Delta_{0}=\sqrt{E_{D}^{2}g^{2}-\varepsilon_{0}^{2}}$. In
the limit $T\gg T_{c}$ , it is also possible to obtain explicit expressions
of $\Delta_{C}$ by expanding the action in powers of $s_{0}$. To
the lowest order in $\delta$, the SPA result $\Delta_{C}\simeq\sqrt{\frac{\delta g\tanh^{2}\left(\frac{\beta\varepsilon_{0}}{2}\right)\coth^{2}\left(\frac{\beta_{c}\varepsilon_{0}}{2}\right)}{\beta\left[1-\tanh\left(\frac{\beta\varepsilon_{0}}{2}\right)\coth\left(\frac{\beta_{c}\varepsilon_{0}}{2}\right)\right]}}$
is recovered. Higher order terms include deviations from SPA due to
quantum fluctuations. 

Natural extensions of this work include the calculation of thermodynamical
observables, odd-even effects, magnetic susceptibilities and the differential
conductance, the outcome of STM experiments. It would also be interesting
to study in detail a hemispherical particle as this geometry can be
investigated experimentally \cite{nmat}. For the sake of clearness
we have focused here in developing the formalism. Some of these applications
will be discussed in future publications \cite{pedro2}.\\
We have studied the combined effect of thermal and quantum fluctuations
in a zero dimensional superconductor. For the first time we have obtained
explicit expressions for $Z$ and $\Delta_{C}(T)$ valid for all temperatures
and to leading order in $\delta/\Delta_{0}$. For intermediate temperatures
both fluctuations contribute substantially to $\Delta_{C}(T)$. These
results provide a solid theoretical framework to describe quantitatively
pairing in confined geometries at finite temperature beyond the mean
field approximation. A problem of current interest in condensed matter,
nuclear and cold atom physics. 
\begin{acknowledgments}
A.M.G. acknowledges financial support from PTDC/FIS/111348/2009, a
Marie Curie International Reintegration Grant PIRG07-GA-2010-26817
and EPSRC grant EP/I004637/1. \end{acknowledgments}

\end{document}